\documentclass[aps,prd,twocolumn,showpacs,superscriptaddress,groupedaddress,nofootinbib]{revtex4}  
\usepackage{graphicx}  
\usepackage{dcolumn}   

\usepackage{graphicx}
\usepackage{amsmath}
\usepackage{amsfonts}
\usepackage{epsfig}
\usepackage{slashed}
\usepackage{braket}
\usepackage{xcolor}
\usepackage{comment}
\usepackage{cases}

\def\bec{\begin{center}}
\def\eec{\end{center}}
\def\beq{\begin{equation}}
\def\eeq{\end{equation}}
\def\bea{\begin{eqnarray}}
\def\eea{\end{eqnarray}}
\def\KD{K\"{a}hler-Dirac }

\def\Phib{\overline{\Phi}}
\def\Psib{\overline{\Psi}}

\begin{document}
\title{Anomalies and symmetric mass generation for \KD fermions}
\author{Nouman Butt}
\affiliation{Department of Physics, University of Illinois at Urbana-Champaign, 1110 W Green St, Urbana, IL 61801} 
\author{Simon Catterall}
\affiliation{Department of Physics, Syracuse University, Syracuse, NY 13244, USA }
\author{Arnab Pradhan}
\affiliation{Department of Physics, Syracuse University, Syracuse, NY 13244, USA }
\author{Goksu Can Toga}
\affiliation{Department of Physics, Syracuse University, Syracuse, NY 13244, USA }

\date{\today}

\begin{abstract}We show that massless \KD fermions exhibit a mixed gravitational anomaly involving an exact $U(1)$ symmetry which is unique to \KD fields. Under this $U(1)$ symmetry
the partition function transforms by a phase 
depending only on the Euler character of the background space. Compactifying
flat space to a sphere we learn that the anomaly vanishes in odd dimensions
but breaks the symmetry down to $Z_4$ in
even dimensions. This $Z_4$
is sufficient to prohibit bilinear terms from arising in the fermionic effective action. Four fermion terms are allowed but require multiples of two flavors of \KD field.
In four dimensional flat space each \KD field can be decomposed into
four Dirac spinors and hence these anomaly constraints
ensure that  eight Dirac fermions or, for real representations,
sixteen Majorana fermions 
are needed for a consistent 
interacting theory. These constraints on fermion number
agree with known results for topological insulators
and recent work on discrete anomalies rooted in the Dai-Freed theorem. Our work 
suggests that \KD fermions may offer an independent path to understanding these
constraints. Finally we point out that this anomaly survives intact 
under discretization and hence is relevant in understanding recent numerical results on
lattice models possessing massive symmetric phases.
\end{abstract}

\maketitle

\section{\label{intro}Introduction}

The \KD equation gives an alternative to the Dirac equation for describing fermions in which
the physical degrees of freedom are carried by antisymmetric tensors rather than
spinors. These tensors
transform under a twisted rotation group that corresponds to the diagonal subgroup
of the usual (Euclidean) Lorentz group and a corresponding flavor symmetry.
In flat space the \KD field can be decomposed into a set of degenerate Dirac spinors but 
this equivalence is lost in a curved background since the coupling to gravity differs
from the Dirac case. Indeed, unlike the case of Dirac fermions, \KD fermions can be defined
on an arbitrary manifold without the need to introduce a frame and spin connection. These
facts were emphasized many years ago by Banks et al. where an attempt was made to identify
the four Dirac fermions residing in each four dimensional \KD field
with the number of generations in the Standard Model \cite{Banks:1982iq}. With precision LEP
data ruling out a fourth generation this idea had to be abandoned. However, in this paper,
we will argue that there may be another natural interpretation for this degeneracy -
it is required in order to write down anomaly free, and hence consistent, theories of
interacting fermions.

These constraints arise because the partition function of a massless free \KD field
in curved space
transforms by a phase under a 
particular global $U(1)$ symmetry with the phase being determined by the
Euler character $\chi$ of the background. The appearance of $\chi$ shows that the
anomaly is gravitational in origin but is distinct from the usual mixed gravitational anomaly
of Weyl fermions \cite{fuji}. If we compactify flat space to a sphere
this anomaly breaks the $U(1)$ to $Z_4$ in even dimensions which is sufficient to prohibit the appearance of fermion bilinear terms in the quantum effective action.
Four fermion terms are however allowed and the simplest operator of this type requires two flavors of \KD fermion. Since a massless \KD field
can be decomposed into two so-called reduced \KD fields, each carrying half the original degrees of freedom, four reduced fields
are required for the minimal four fermion interaction.

In even dimensions a reduced \KD field, in the flat space limit, can be decomposed into $2^{D/2}$ Majorana spinors 
and we learn that consistent interacting theories in even dimensions possess
$2^{D/2+2}$ Majorana fields. These fermion numbers 
agree with a series of anomaly cancellation conditions associated with certain discrete
symmetries in dimensions two and four - see table~\ref{tab1} and  \cite{Garcia-Etxebarria:2018ajm,Wan:2020ynf}.
\begin{table}[h]
    \centering
    \begin{tabular}{||c|c|c||}\hline
       D  &  Symmetry & Critical number of Majoranas\\\hline
        2 & Chiral fermion parity & 8\\\hline
        4 & Spin-$Z_4$ & 16\\\hline
    \end{tabular}
    \caption{Number of Weyl fermions needed for consistent interacting theories in $D=2$ and $D=4$ }
    \label{tab1}
\end{table}
Since cancellation of all 't Hooft anomalies is a necessary condition for
fermions to acquire mass without
breaking symmetries this suggests it may be possible to build models with precisely
this fermion content where all fermions are gapped in the I.R. In fact there are
examples in condensed matter physics where precisely this occurs
see \cite{Fidkowski:2009dba,Ryu:2012he,Kapustin:2014dxa,You:2014vea,Morimoto:2015lua,You:2017ltx,Wang:2020iqc,Guo:2018vij}. Similar results have been obtained in
staggered fermions in lattice gauge theory \cite{Ayyar:2014eua,Ayyar:2015lrd,Ayyar:2016lxq,Catterall:2015zua,Catterall:2017ogi,Catterall:2018lkj}.

The plan of the paper is as follows. We start in section~\ref{KD} by giving a brief introduction to \KD fermions exhibiting
their connection to Dirac fermions, and showing how, in the case of massless fields, the
theory is invariant under a $U(1)$ symmetry whose generator $\Gamma$ anticommutes with
the \KD operator on any curved background. Using
$\Gamma$ one can project out half the degrees of freedom to obtain
a {\it reduced} \KD field. Section~\ref{anomaly} shows that the $U(1)$ symmetry suffers
from a gravitational anomaly in even dimensions and breaks to $Z_4$.
In section~\ref{spectral}
we show using a spectral flow argument that this remaining $Z_4$ symmetry suffers from
a global anomaly in the presence of interactions unless the theory contains multiples of four reduced \KD fields.
In section~\ref{SMG} we point out that a necessary
condition for symmetric mass generation in such theories is that these 
anomalies cancel and we give examples of possible four fermion interactions that
might be capable of achieving this. Section~\ref{lattice} points out the connections between
these continuum ideas and lattice fermions and a final summary appears in
section~\ref{end}.

\section{\label{KD}Review of K\"{a}hler-Dirac Fields}
The \KD equation 
arises on taking the square root of the Laplacian operator written in terms of exterior derivatives. We start by defining the \KD operator
$K=d-d^\dagger$ with $d^\dagger$ the adjoint of $d$. 
Clearly $-K^2=dd^\dagger+d^\dagger d=\Box$ since the derivative operators are nilpotent.
This suggests an alternative to the usual Dirac equation \cite{kahler}:
\beq
(K-m)\Phi=0\label{KDeq}\eeq
where $\Phi=\left(\phi_0,\phi_1,\ldots \phi_p,\phi_D\right)$ 
is a collection of $p$-forms. 
The action of the derivative operators on these forms is then given by \cite{Banks:1982iq,Catterall:2018lkj}
\begin{align}
\begin{split}
    d\Phi=\Big(0,\partial_\mu\phi, \partial_{\mu_1}\phi_{\mu_2}-\partial_{\mu_2}\phi_{\mu_1},\ldots\\
    \qquad,\sum_{perms\, \pi}\left(-1\right)^\pi\partial_{\mu_1}\phi_{\mu_2\ldots\mu_D}\Big)
\end{split}
\end{align}
\beq
-d^\dagger\Phi=\left(\phi^\nu,\phi^\nu_\mu,\ldots ,\phi^\nu_{\mu_1,\ldots ,\mu_{D-1}},0\right)_{;\nu}.\eeq
An inner product of two such \KD fields $A$ and $B$ can be defined as
\begin{equation}
    \left[A,B\right]=\sum_p \frac{1}{p!}a^{{\mu_1}\ldots{\mu_D}}b_{{\mu_1}\ldots{\mu_D}}.
\end{equation}
Using this allows one to obtain the \KD equation by the variation of the \KD action
\begin{equation}
    S_{\rm KD}=\int d^Dx\sqrt{g}\,\left[\Phib ,\left( K-m\right)\Phi\right]
\end{equation}
where $\Phib$ is an independent (in Euclidean space) \KD field.

It is easy to see
that the \KD operator anticommutes
with a linear operator $\Gamma$ which acts on the $p$-form fields as
\beq\Gamma:\quad \phi_p\to \left(-1\right)^p\phi_p.\label{U1trans}\eeq  
This anti-commutation property can be used
to construct a $U(1)$ symmetry of the massless \KD action which acts on $\Phi$ as
\begin{equation}
    \Phi\to e^{i\alpha\Gamma}\Phi.
\end{equation}
Furthermore, using $\Gamma$ one can define operators which project out even and odd form
fermions - so-called reduced \KD fields $\Phi_\pm=P_\pm\Phi$ with
\beq
P_\pm=\frac{1}{2}\left(1\pm \Gamma\right).\eeq
The \KD operator $K$ couples even to odd forms and hence
the massless
\KD action separates 
into two independent pieces $S=\int \left(\Phib_+\,K\Phi_-+\Phib_-\,K\Phi_+\right)$.
Retaining just one of these terms 
one obtains an action for such a reduced \KD (RKD) field 
\beq
S_{\rm RKD}=\int d^Dx\sqrt{g}\,\left[\Phib_- ,K \Phi_+\right].
\end{equation}
Notice that the single flavor reduced theory admits no
mass term since $\Phib_+\Phi_-=0$.
Finally if we relabel $\Phib_-\to \Phi_-$ 
this reduced action can be rewritten in a Majorana-like form
\beq
S_{\rm RKD}=\frac{1}{2}\int d^Dx\sqrt{g}\,\left[\Phi ,K\Phi\right].\label{RKD}\eeq

Given that both the Dirac operator and the \KD operator correspond to square
roots of the Laplacian one might imagine that there is a relation between the \KD field
and spinor fields.
To exhibit this relationship we construct a matrix $\Psi$ by combining
the p-form components of the \KD field $\Phi$ with products of Dirac gamma matrices 
\beq
\Psi=\sum_{p=0}^D\frac{1}{p!} \gamma^{{\mu_1}\ldots{\mu_p}}\phi_{{\mu_1}\ldots{\mu_p}}.
\label{KDtomatrix}\eeq
where $\gamma^{{\mu_1}\ldots{\mu_p}}=\gamma^{\mu_1}\gamma^{\mu_2}\cdots\gamma^{\mu_p}$ are constructed using the
usual (Euclidean) Dirac matrices $\gamma^\mu=\gamma^a e^\mu_a$~\footnote{We will restrict ourselves to even dimensions in what follows. Odd dimensions require twice as many spinor degrees of freedom to match the number of components of a \KD field - see \cite{Hands:2021mrg}}.
In flat space it is straightforward to show that the matrix $\Psi$ 
satisfies the usual Dirac equation 
\beq
\left(\gamma^\mu\partial_\mu -m\right)\Psi=0\eeq
and describes $2^{D/2}$ degenerate Dirac spinors corresponding to the columns
of $\Psi$. This equation of motion can be derived from
the action
\begin{equation}
    S=\int d^Dx\, {\rm Tr}\,\left[\Psib(\gamma^\mu \partial_\mu-m)\Psi\right].
\end{equation}
This action is invariant under a global ${\rm Spin}(D)\times SU(2^{D/2})$ 
symmetry where the first factor corresponds to Euclidean Lorentz transformations
and the second to an
internal flavor symmetry. In
even dimensions we can write a matrix representation of the $U(1)$ generator as
$\Gamma\equiv \gamma_5\otimes \gamma_5$ where the two factors
act by left and right multiplication on $\Psi$.
The matrix representation of a reduced \KD field is then given by
\begin{equation}
\Psi_\pm =\frac{1}{2}\left(\Psi\pm\gamma_5\Psi\gamma_5\right).
\end{equation}
Similarly the reduced action can be written as 
\begin{equation}
\int d^Dx\,{\rm Tr}\,\left(\overline{\Psi}_-\gamma^\mu\partial_\mu\Psi_+\right).
\end{equation}
The condition $\Phib_-=\Phi_-$ then implies the matrix condition
\begin{equation}\Psi^*=B\Psi B^T\label{reality}\end{equation}
where $B=C\gamma_5$ with $C$ the usual
charge conjugation matrix. Using this one can write a Majorana-like matrix representation of
the reduced action as
\beq \frac{1}{2}\int d^Dx\,{\rm Tr}\,\left(B\Psi^TB^T\gamma^\mu\partial_\mu \Psi\right).\eeq
Notice that after this reduction the free theory in 
flat space corresponds to $2^{D/2-1}$ Dirac or 
$2^{D/2}$ Majorana spinors.

We can gain further insight into the
reality condition eqn.~\ref{reality} by going to a 
chiral basis for the gamma matrices. In four dimensions 
the full \KD field $\Psi$ then takes the form
\begin{equation}
\Psi=\left(\begin{array}{cc}E&O^\prime\\
O&E^\prime\end{array}\right)\label{blocks}
\end{equation}
where $O$ and $O^\prime$ denote $2\times 2$ blocks of odd form fields while $E$ and $E^\prime$ denote corresponding even form fields. Each block contains a doublet
of Weyl fields which transform in representations of the $SU(2)_L\times SU(2)_R$ Lorentz
and an $SU(2)\times SU(2)$ flavor symmetry. The condition 
$\Psi^*=B\Psi B^T$ implies $O^\prime=i\sigma_2 O^* i\sigma_2$ and $E^\prime=-i\sigma_2 E^*i\sigma_2$. 
This suggests that the operation $X\to  i\sigma_2 X^* i\sigma_2$ can be interpreted
as a generalized
charge conjugation operator that flips both chirality and flavor representation of
a given Weyl doublet within the \KD field. It also implies that both $(O,O^\prime)$ and $(E,E^\prime)$ constitute doublets of Majorana spinors. 

Finally we should note that while the \KD equation eqn.~\ref{KDeq} written
in the language of forms does not change in curved space, its
matrix representation takes the modified form
\begin{equation}
    (e^\mu_a\gamma^a D_\mu-m)\Psi=0
\end{equation}
where $e_\mu^a$ is the vielbein or frame and $D_\mu\Psi=\partial_\mu\Psi+[\omega_\mu,\Psi]$ is the covariant
derivative associated to the spin connection. 

\section{\label{anomaly}A gravitational anomaly for K\"{a}hler-Dirac fields}
In the previous section we showed that the \KD action is invariant under a $U(1)$ symmetry.
However in the quantum theory we should also be careful to examine the invariance of
the fermion measure. We will do this for a generic four dimensional curved background using
the matrix representation of the \KD theory. The curved space action in $D=4$ reads 
\beq
S=\int d^4x \sqrt{g}\, {\rm Tr}\,\left(\Psib e^\mu_a\gamma^a D_\mu\Psi\right)\eeq
where $D_\mu\Psi=\partial_\mu\Psi+[\omega_\mu,\Psi]$ with $\omega_\mu$
the spin connection and we have introduced the frame field $e_\mu^a(x)$ to translate between
flat and curved space indices with $e_\mu^a e_\nu^b\delta_{ab}=g_{\mu\nu}$.
In the standard way we start by expanding 
$\Psi$ and $\Psib$ on a basis of eigenstates of the \KD operator:
\beq
\gamma^\mu D_\mu \phi_n=\lambda_n \phi_n\eeq
with $\int d^4x\, e(x){\rm Tr}\left(\overline{\phi}_n(x)\phi_m(x)\right)=\delta_{nm}$ and $e={\rm det}\,(e_\mu^a)$.
Thus
\begin{align}
\Psi(x)&=\sum_n a_n\phi_n(x)\\
\Psib(x)&=\sum_n \overline{b}_n\overline{\phi}_n(x).
\end{align}
The measure is then written $D\Psib\,D\Psi=\prod_n d\overline{b}_n\, da_n$ and the variation of
this measure under the $U(1)$ transformation with parameter $\alpha(x)$ is given by $e^{-2i\int \alpha(x)A(x)}$ where the anomaly $A$ is formally given by
\beq A(x)={\rm Tr}\,\sum_n e\overline{\phi}_n \Gamma\phi_n(x)\label{index}\eeq
where the operator $\Gamma=\Gamma_5\otimes \gamma_5$ carries a 
flavor rotation matrix $\gamma_5$ acting on the right of the matrix field 
and a curved space chiral matrix $\Gamma_5$ acting on
the left with 
$\Gamma^5=\gamma^a\gamma^b\gamma^c\gamma^d e^1_ae^2_be^3_ce^4_d=\gamma^5e$.
We need a gauge invariant regulator for this expression so we try inserting the factor
\beq e^{\frac{1}{M^2}\left(\gamma^\mu D_\mu\right)^2}\eeq
into the expression for $A$.
We can write
\begin{equation}
\left(\gamma^\mu D_\mu\right)^2=D^\mu D_\mu+e^\mu_c e^\nu_d \sigma^{cd}[D_\mu, D_\nu]
\label{dslashsq}
\end{equation}
where $\sigma^{cd}=\frac{1}{4}[\gamma^c,\gamma^d]$ are the generators of ${\rm Spin}(4)$.
Furthermore for KD fermions we have:
\[[D_\mu,D_\nu]\psi=[R_{\mu\nu},\psi]\] 
where $R_{\mu\nu}=\frac{1}{2}R_{\mu\nu}^{ab}\sigma_{ab}$. 
Plugging this expression into eqn.~\ref{dslashsq} yields
\[ \left(\gamma^\mu D_\mu\right)^2\psi= D^\mu D_\mu\psi+
{\frac{1}{2}}e^\mu_c e^\nu_d \sigma^{cd}R_{\mu\nu}^{ab}
[\sigma_{ab},\psi].\]
The anomaly can then be written
\begin{align}
A(x)&=\lim_{M\to\infty}{\rm Tr}\,\sum_n e\left(\overline{\phi}_n{\Gamma }e^{\frac{1}{M^2}(\gamma^\mu D_\mu)^2}\phi_n\right)\\\nonumber
&=\lim_{M\to\infty}{\rm Tr}\, \left(\Gamma e^{\frac{1}{M^2}(\gamma^\mu D_\mu)^2}\sum_n e \phi_n \overline{\phi}_n\right)\\\nonumber
&=\lim_{x\to x^{'}}\lim_{M\to\infty}{\rm Tr}\, \left(\Gamma e^{\frac{1}{M^2}(\gamma^\mu D_\mu)^2}\delta(x - x^{'})\right)\\\nonumber
\begin{split}
    &=\lim_{x\to x^{'}}\lim_{M\to\infty}{\rm Tr}\,\Big(e\gamma^5 e^{\frac{1}{M^2}(D^\mu D_\mu + {\frac{1}{2}}e^\mu_c e^\nu_d {\sigma^{cd}}R_{\mu\nu}^{ab}[\sigma_{ab},.])}\\
    &\qquad \times \delta(x - x^{'})\gamma_5\Big). \nonumber
\end{split}
\end{align}
Expanding the exponential to $O(1/M^4)$ to get a non-zero result for the trace over spinor
and flavor indices  and acting with $e^{\frac{1}{M^2}D_\mu D^\mu}$ on the delta
function yields~\footnote{See appendix for more details}
\begin{align}
\begin{split}
    A &=\frac{1}{16\pi^2}\left(\frac{1}{2!}\right)\left(\frac{1}{4}\right) {\rm tr}\,
\left(e\gamma^5\sigma^{ab}\sigma^{cd}\right)e^\mu_ae^\nu_b e^\rho_c e^\lambda_d R_{\mu\nu}^{CD}R_{\rho\lambda}^{EF}\\
&\qquad \times {\rm tr}\,\left(\sigma_{CD}\sigma_{EF}\gamma_5\right) \nonumber
\end{split}
\\
&= \frac{1}{128\pi^2}\epsilon^{\mu\nu\rho\lambda}\epsilon_{CDEF}
R_{\mu\nu}^{CD}
R_{\rho\lambda}^{EF}\label{anomaly2}
\end{align}
where we have also employed the result:
\[e\epsilon^{abcd}e^\mu_a e^\nu_b e^\rho_c e^\lambda_d=\epsilon^{\mu\nu\rho\lambda}.\]
Thus the anomaly $A(x)$ is just the Euler density and we find that
the phase transformation of the partition function under the global $U(1)$ symmetry is then
\beq
Z\to e^{-2i\alpha\int d^4x\, A(x)}=e^{-2i\alpha\chi}Z.\eeq
This result for the anomaly agrees with a previous lattice calculation that employed a discretization of the \KD action on simplicial lattices \cite{Catterall:2018lkj}.
The non-zero value for the anomaly originates in the
existence of exact zero modes of the \KD operator. Such zero modes are eigenstates of $\Gamma$ 
and eqn.~\ref{index} shows that $\int d^4x\, A(x)=n_+-n_-$ where $n_\pm$ denotes the number of zero
modes with $\Gamma=\pm 1$. Our final result is then a consequence of the index theorem
\beq n_+-n_-=\chi.\eeq
On the sphere (which can be regarded as a compactification of $R^4$) the presence of this phase breaks
the $U(1)$ to $Z_4$. 
This non-anomalous $Z_4$ is then sufficient to prohibit fermion bilinear mass terms from
appearing in the effective action of the theory. Four fermion terms are allowed 
but require at least two flavors of \KD field to be non-vanishing~\footnote{Of course we
can go further and demand that the anomaly be cancelled for
manifolds with other values of $\chi$. For example, on manifolds with $\chi=-4$ only
eight fermion terms are allowed, only twelve 
fermion terms on spaces with $\chi=-6$ etc. Other work on higher order multifermion interactions
can be found in \cite{Wu:2019isp,Jian:2019zxu}.}.
Since each massless 
\KD field can be written in terms of two independent reduced fields this implies consistent,
interacting theories require four reduced \KD fields. In four dimensions, and taking
the flat space limit, each such reduced field corresponds to four Majorana fermions 
and we learn that such theories contain sixteen Majorana fermions.
It is not hard to generalize this argument to any even dimension.

\section{\label{spectral}A global $Z_4$ anomaly}

In the last section we found that a system of free \KD fermions propagating on an
even dimensional space is anomalous and remains invariant only under
a $Z_4$ symmetry. In this section we will examine such a theory in the
presence of interactions
and show that this residual $Z_4$ symmetry suffers from a global anomaly
unless the theory contains multiples of four reduced \KD fields. 

From eqn.~\ref{RKD} it is clear that the
effective action for a reduced \KD fermion is given by a Pfaffian ${\rm Pf}(K)$.
From the property $[\Gamma,K]_+=0$ it is easy to show that ${\rm det}\,(K)\ge 0$. However since
${\rm Pf}\,(K)=\pm \sqrt{\rm det}\,(K)$ 
there is an ambiguity in the phase of the Pfaffian.

To analyze this in more detail we
will consider the theory on the non-orientable manifold $RP^4$ 
and deform the theory to remove the one zero mode. The simplest possibility is to couple
a pair of such reduced \KD fields to an auxiliary real scalar field $\sigma$. The fermion
operator is then given by
\beq
M=\delta^{ab}K+\sigma(x)\epsilon^{ab}.\eeq
We will assume that the total action (including terms involving just $\sigma$)
is invariant under a discrete symmetry which extends the fermionic $Z_4$ discussed
in the previous section:
\begin{align}
    \Phi&\to i\Gamma \Phi\\
    \sigma&\to -\sigma.
\end{align}
Notice that this fermion operator is antisymmetric and real and hence all eigenvalues
of $M$ lie on the imaginary axis. Let us define the Pfaffian as the product of the eigenvalues
in the upper half plane in the background of some reference configuration
$\sigma=\sigma_0={\rm constant}$. By continuity we define the Pfaffian to be the product of these same eigenvalues under fluctuations of $\sigma$. Furthermore, it is easy to see that
as a consequence of the $Z_4$ symmetry
\beq 
\Gamma M\left(\sigma\right)\Gamma=-M\left(-\sigma\right)\label{pfaff}.\eeq
This result shows that the spectrum and hence the determinant is indeed invariant under 
the $Z_4$ transformation $\sigma\to -\sigma$. But this is not enough to show the Pfaffian itself is unchanged since 
there remains the possibility that eigenvalues flow through the origin as $\sigma$ is deformed smoothly to $-\sigma$ leading to a sign change.
To understand what happens we consider a smooth interpolation of $\sigma$:
\beq 
\sigma(s)=s\sigma_0\quad s\in \left(-1,+1\right).\eeq
The question of eigenvalue flow can be decided by focusing on the behavior of
the eigenvalues of the fermion operator closest to the origin at small $s$.
In this region the eigenvalues of smallest magnitude
correspond to fields which are constant over the lattice and satisfy the 
eigenvalue equation:
\beq 
\sigma_0 s\epsilon^{ab}v^b=\lambda v^a.\eeq
The two eigenvalues $\lambda=\pm i\sigma_0 s$. Clearly these eigenvalues change sign as
as $s$ varies from positive to negative values leading to a Pfaffian sign change. 
This can also be seen
explicitly from eqn.~\ref{pfaff} since 
\beq 
{\rm Pf}\,\left[M(-\sigma)\right]={\rm det}\,\left[\Gamma\right]{\rm Pf}\,\left[M(\sigma)\right]=-{\rm Pf}\,\left[M(\sigma)\right].\eeq
We thus learn that the Pfaffian of the 2 flavor
system indeed changes sign under the $Z_4$ transformation. On integration
over $\sigma$ the 
value of any even function of $\sigma$ including the partition
function would then yield zero rendering expectation values of $Z_4$ invariant operators
ill-defined. This corresponds to a mixed global 't Hooft anomaly between the discrete $Z_4$ symmetry and a gauged $Z_2$ reflection symmetry
in the interacting theory.

Clearly this global anomaly will be cancelled
for any multiple of four reduced \KD fields provided that
they couple via a Yukawa term of the form $\sigma(x)\Phi^a C^{ab}\Phi^b$ with $C$ a real,
antisymmetric matrix. In that case $C$ can be brought to a canonical antisymmetric form $\left(\lambda_1i\sigma_2\otimes\lambda_2i\sigma_2\otimes ...\right)$ using a non-anomalous orthogonal transformation. Positivity of the Pfaffian 
under $\sigma\to -\sigma$ then depends on the Yukawa interaction in $M$
containing an even number of such $2\times 2$ blocks.
Decomposing the reduced \KD field in a flat background into
spinors we see that anomaly cancellation occurs for
eight or sixteen Majorana fermions in two and four dimensions respectively. 

The spectral
flow argument we have given is similar to the one given by 
Witten in showing that a single Weyl
fermion in the fundamental representation of $SU(2)$ is anomalous \cite{Witten:1982fp}.

\section{\label{SMG}Symmetric mass generation}

The cancellation of anomalies is crucial to the problem of giving masses to fermions
without breaking symmetries. Since anomalies originate from massless states then any phase where all states are massive in the I.R must necessarily arise from a U.V theory with vanishing
anomaly. In particular, it is only possible to accomplish such symmetric
mass generation if one cancels
off the 't Hooft anomalies 
for all global symmetries \cite{Razamat:2020kyf,Tong:2021phe}. 

In the previous
section we have seen that only 
multiples of four reduced \KD fields have vanishing $Z_4$ anomaly. Thus we
require that any interactions we introduce in the theory respect this symmetry. The
simplest such interaction is a four fermion operator as we have already discussed. It
corresponds to adding a simple $\int \sigma^2$ term to the Yukawa action discussed in the
previous section. 

We should note however that cancellation of anomalies is a necessary condition for symmetric mass generation but it may not be sufficient -- the fact that four fermion terms are perturbatively irrelevant operators in dimensions greater than two may mean that a more complicated
scalar action is required -- indeed this was the finding of numerical work in four
dimensions where a continuous phase transition to a massive symmetric phase was found only by tuning an additional scalar kinetic term \cite{Butt:2018nkn}.

With this caveat it is useful to give examples of possible four fermion terms that might lead to symmetric mass generation
For example, one can imagine taking four reduced \KD fields transforming in
the fundamental representation of a $SO(4)$ symmetry and employ the term
\beq
\int d^Dx\, \sqrt{g}\,\left(\left[\Phi^a,\Phi^b\right]_+\right)^2\eeq
where the $+$ subscript indicates that fermion bilinear is projected to the self-dual $(1,0)$ representation of $SO(4)$. 
In practice this can be implemented in flat space via a Yukawa term which is given in the matrix representation by
\beq
\int d^Dx\, G\,{\rm Tr}\,\left(\Psi^a(x)\Psi^b(x)\right)_+\sigma_{ab}(x)+\frac{1}{2}\sigma^2_{ab}(x).\label{ffermion}\eeq
Notice that this Yukawa interaction is mediated by a scalar $\sigma_{ab}^+$ that also
transforms in the self-dual representation of $SO(4)$
\begin{equation}
    \sigma_{ab}^+=\frac{1}{2}\left(\sigma_{ab}+\frac{1}{2}\epsilon_{abcd}\sigma_{cd}\right).
\end{equation}

In the next section we show how \KD theories can be discretized in a manner
which leaves the anomaly structure of the theory intact and results in theories
of (reduced) staggered fermions. The four fermion interaction that results from
integrating over $\sigma_+$ in eqn.~\ref{ffermion} has been studied using numerical simulation and 
the results of this work indeed provide evidence of a massive
symmetric phase in dimensions two and three 
\cite{Ayyar:2014eua,Ayyar:2015lrd,Ayyar:2016lxq,Catterall:2015zua} while
an additional scalar kinetic operator was also needed in four dimensions \cite{Butt:2018nkn}.

As another example one can take eight flavors of reduced \KD field
which are taken to transform in the eight dimensional real 
spinor representation of ${\rm Spin}(7)$.
An appropriate Yukawa term which might be used to 
gap those fermions is given by
\beq 
\int d^Dx\, {\rm Tr}\,\left(\Psi^a(x)\Gamma^{ab}_\mu\Psi^b(x)\right)\sigma_\mu(x)\eeq
where $\Gamma_\mu,\mu=1\ldots 7$ are the (real) Dirac matrices for ${\rm Spin}(7)$ \cite{You:2014vea}. This
interaction was shown by Fidkowski and 
Kitaev to gap out boundary Majorana modes in a (1+1)-dimensional system
without breaking symmetries \cite{Fidkowski:2009dba}. This interaction may also play a role
in constructing \KD theories that target GUT models. For example, if one is able to gap
out the $(E,E^\prime)$ blocks occurring in eqn.~\ref{blocks} for a reduced \KD field valued in ${\rm Spin}(7)$ the remaining light fields live in the representation $(8,2,1)$. If the
${\rm Spin}(7)$ is subsequently Higgsed to ${\rm Spin}(6)=SU(4)$ then this representation
breaks to $(4,2,1)\oplus(\overline{4},1,2)$ which is the field content of the Pati-Salam
theory \cite{Pati}.

\section{\label{lattice}Exact anomalies for lattice fermions}

One of the most important properties of the \KD equation is that it can be discretized
without encountering fermion doubling \cite{Rabin:1981qj, Becher:1982ud}. Furthermore this discretization procedure can be
done for any random triangulation of the space allowing one to capture topological
features of the spectrum. The idea is to replace continuum p-forms by p-cochains or
lattice fields living on oriented p-simplices in the triangulation. The exterior derivative and
its adjoint are mapped to co-boundary and boundary operators which act naturally on
these p-simplices and retain much of the structure of their continuum cousins -- for
example they are both nilpotent operators. Homology theory can then be used to
show that the spectrum of the discrete theory evolves smoothly into its continuum cousin
as the lattice spacing is sent to zero -- there are no additional lattice modes or doublers that obscure the continuum limit. Furthermore, the number of exact zero modes of the lattice \KD operator is exactly the same as found in the continuum theory. This
immediately suggests that the anomaly encountered earlier which depends only on the
topology of the background space can be exactly reproduced in the lattice theory. This
was confirmed in \cite{Catterall:2018lkj} where a lattice calculation revealed
precisely the same gravitational anomaly derived in this paper.

If one restricts to regular lattices with the topology of the torus it is straightforward to see that the discrete \KD operator discussed above can be mapped
to a staggered lattice fermion operator on a lattice with half the lattice spacing
\cite{Banks:1982iq}. One simply maps the p-form components located in the
matrix $\Psi^a$ into a set of single
component lattice fermions $\chi^a$ via
\beq  
\Psi(x)=\sum_{n_\mu=0,1}\chi(x+n_\mu)\gamma^{\left(x+n_\mu\right)}\eeq
where $\gamma^x=\prod_{i=1}^D\gamma_i^{x_i}$ and the summation runs over the $2^D$ points
in a unit hypercube of a regular lattice. If one substitutes this expression into
the continuum kinetic term, replaces the continuum derivative with a symmetric finite difference and carries out the trace operation one obtains the free staggered fermion action.
Indeed the operator $\Gamma$ acting on forms then becomes the site parity operator $\epsilon(x)=\left(-1\right)^{\sum_{i=1}^Dx_i}$ and the $U(1)$ symmetry of the massless
\KD action is just the familiar
$U(1)_\epsilon$ symmetry of staggered fermions. Indeed, it is possible to repeat the arguments
of section~\ref{spectral} to show that a staggered fermion theory equipped with
a four fermion term is only well-defined for multiples of four reduced staggered
fermions under which the classical $Z_4$ symmetry is preserved. This helps to
explain why these theories seem capable of generating a massive symmetric phase
\cite{Ayyar:2014eua,Ayyar:2015lrd,Ayyar:2016lxq,Catterall:2015zua,Butt:2018nkn}.

\section{\label{end}Summary}
In this paper we have shown that theories of 
massless \KD fermions suffer from a gravitational anomaly that breaks a $U(1)$ symmetry down to $Z_4$ in even dimensions. We derive this anomaly by 
computing the symmetry variation of the path integral for free \KD fermions propagating in 
a background curved space. The remaining $Z_4$ prohibits fermion bilinear mass terms from
arising in the quantum effective action.  We then use spectral flow arguments
to argue that multiples of four flavors of \KD are needed to 
avoid a further global anomaly in this $Z_4$ symmetry in the presence of interactions. Since
four fermion interactions are allowed by these constraints we argue that they
may be capable of gapping such systems without breaking symmetries.

In flat space each reduced \KD field transforming in a real representation 
can be decomposed into $2^{D/2}$ Majorana fermions.
Thus anomaly cancellation in the interacting theory dictates a very specific 
fermion content - multiples of eight
and sixteen Majorana
fermions in two and four dimensions respectively. Remarkably,
this fermion counting agrees with independent constraints based on the cancellation of 
the chiral fermion parity and spin-$Z_4$ symmetries
of Weyl fermions in two and four dimensions \cite{Garcia-Etxebarria:2018ajm,Wang:2020iqc}~\footnote{
One can also decompose a \KD fermion
into two and four Majorana spinors in one and three dimensions respectively. Building
four fermion operators for these \KD fields then yields theories with eight and
sixteen Majorana spinors which is also in agreement 
with results from odd dimensional topological insulators.}.

Finally we discuss how this anomaly can be realized exactly in lattice realizations of
such systems and emphasize how the results in this paper shed light on the appearance of
massive symmetric phases in recent simulations of lattice four fermion models. The appearance of an anomaly in a lattice system  is
notable as it contradicts the usual folklore that anomalies only appear in systems with an
infinite number of degrees of freedom.

While the anomaly vanishes for closed odd dimensional manifolds it is non-zero for
odd dimensional manifolds with boundary. For example  the Euler
characteristic of the three ball is $\chi(B^3)=1$ and the
symmetry in the bulk is hence broken to $Z_2$ allowing for the presence of mass terms. 
However the
boundary fields living on $S^2$ possess an enhanced $Z_4$ symmetry prohibiting
such bilinear terms and we learn that such boundary fields can instead be gapped using four fermion interactions.

\acknowledgments
This work was supported by the US Department of Energy (DOE), Office of Science, Office of High Energy Physics under Award Number {DE-SC0009998}. SC is grateful for helpful
discussions with Erich Poppitz, David Tong and Yuven Wang.

\appendix 
\section{Delta function}
Following \cite{fuji}

\[\delta(x - x^{'}) = \int \frac{d^4 k}{(2\pi)^4} e^{ik_\mu D^\mu \sigma(x,x^{'})} \]
where $D_\mu$ is now the \KD operator and $\sigma(x,x^{'})$ is the geodesic biscalar [a generalization of $\frac{1}{2}(x - x^{'})^2$ in flat space] defined by
\[\sigma(x,x^{'}) = \frac{1}{2}g^{\mu\nu}D_\mu \sigma (x,x^{'}) D_\nu \sigma (x,x^{'})\]
with
\[\sigma (x,x) = 0\]
and
\[ \lim_{x\to x^{'}} D_\mu D^\nu \sigma (x,x^{'}) = g^\nu_\mu \]\label{identity}

Now,
\begin{align}
\begin{split}
D^2 \delta(x - x^{'}) &= D^\nu \int \frac{d^4 k}{(2\pi)^4} [ik_\lambda D_\nu D^\lambda \sigma \\ 
&\qquad+ \frac{1}{2!}\{(D_{\nu}(ik\cdot D\sigma))(ik\cdot D\sigma) \\
&\quad\qquad +(ik\cdot D\sigma)(D_{\nu}(ik\cdot D\sigma)\} + ... ]
\end{split}
\\
\begin{split}
&= \int \frac{d^4 k}{(2\pi)^4}[D^\nu(ik_\lambda D_\nu D^\lambda \sigma) \\
&\qquad + (ik_\lambda D_\nu D^\lambda \sigma)(ik_\rho D^\nu D^\rho \sigma) + ... ].
\end{split}
\end{align}
Taking $\lim_{x\to x^{'}}$, other terms represented by ... vanish, and we obtain
\begin{align}
D^2 \delta(x - x^{'}) &= \int \frac{d^4 k}{(2\pi)^4}[(ik_\lambda g_\nu^\lambda)(ik_\rho g^{\nu\rho}) + D^\nu(ik_\lambda g_\nu^\lambda)]\\
&= \int \frac{d^4 k}{(2\pi)^4}[(ik_\nu)(ik^\nu) + iD^\nu(k_\nu)]\\
&= \int \frac{d^4 k}{(2\pi)^4}[-k^2 + iD^\nu(k_\nu)].
\end{align}
Which implies
\begin{equation}
    \lim_{x\to x^{'}} e^{D^2}\delta(x - x^{'}) = 
\int \frac{d^4 k}{(2\pi)^4}e^{[-k^2 + iD^\nu(k_\nu)]}.
\end{equation}
Hence,
\begin{align}
\lim_{x\to x^{'}} e^{D^2/M^2}\delta(x - x^{'}) &= \int \frac{d^4 k}{(2\pi)^4}e^{[-k^2 + iD^\nu(k_\nu)]/M^2}\\
&= M^4 \int\frac{d^4 k}{(2\pi)^4}e^{[-k^2 + iD^\nu(k_\nu)/M]}\\
&= M^4\frac{1}{16 \pi^2}.\label{deltafunc}
\end{align}
\bibliography{chiralKD}

\end{document}